%% file: big2017.tex
\UseRawInputEncoding

\documentclass[conference]{IEEEtran}

\usepackage[dvips]{graphicx}

\ifCLASSINFOpdf
\else
\fi
\begin{document}
%
\title{Experiments of posture estimation on vehicles using wearable acceleration sensors}

\author{\IEEEauthorblockN{Yoji Yamato}
\IEEEauthorblockA{Software Innovation Center, NTT Corporation\\
Musashino-shi, Tokyo 180--8585, Japan\\
Email: yamato.yoji@lab.ntt.co.jp}}


%


\maketitle

\begin{abstract}
In this paper, we study methods to estimate drivers' posture in vehicles using acceleration data of wearable sensor and conduct a field test. Recently, sensor technologies have been progressed. Solutions of safety management to analyze vital data acquired from wearable sensor and judge work status are proposed. To prevent huge accidents, demands for safety management of bus and taxi are high. However, acceleration of vehicles is added to wearable sensor in vehicles, and there is no guarantee to estimate drivers' posture accurately. Therefore, in this paper, we study methods to estimate driving posture using acceleration data acquired from T-shirt type wearable sensor hitoe, conduct field tests and implement a sample application.
\end{abstract}

%
\IEEEpeerreviewmaketitle

\input{section1E}
\input{section2E}
\input{section3E}
\input{section4E}
\input{section5E}
\input{section6E}

\section*{Acknowledgment}
The author would like to thank Hideki Hayashi who was a manager of this research.



%

\input{reference}
\end{document}

%% file: section1E.tex
\section{Introduction}
One of IoT application areas (e.g., \cite{JIP3}\cite{IEEJ2}), there are solutions of safety management which acquire users' vital data by wearable sensor and analyze health and work status of them. Sensors acquire vital data such as acceleration and heart rate, and the data is managed and analyzed using cloud technologies \cite{IEEE}-\cite{IEICE2}. A cloud manages each user's data securely (e.g., \cite{TC}\cite{ACM}), analyzes vital data using processing of batch, micro-batch like Spark Streaming \cite{Spark} and stream like Storm \cite{Storm} and judges users' statuses such as postures of falling down, fatigue level change promptly. And when there are problems of statuses, a cloud conducts actions such as staff assignment and emergent alert by coordinating outer systems using Web services or other coordination technologies \cite{SAINT}\cite{ICWS2}. 

In Japan, a fatal accident of long distance bus occurred in which 15 people were dead in January 2016 \cite{Ski}. Thus, needs for safety management with wearable sensors for drivers are increasing because dangerous driving posture like picking up things or fatigue accumulation of bus or taxi (hereafter, vehicle) drivers may result in accidents. However, acceleration of vehicles is added to acceleration of wearable sensor in vehicles, and there is no guarantee to estimate drivers' posture accurately. Therefore, in this paper, we study methods to estimate driving posture using acceleration data acquired from T-shirt type wearable sensor hitoe\cite{hitoe} and conduct field tests.

%% file: section2E.tex
\section{Existing wearable sensor technologies}
Regarding to sensors for acquiring vital data, wearable terminals have been spread. There are various terminals such as watch type, list band type, eyeglass type, T-shirt type and so on. Apple Watch\cite{Apple} is a watch type computer, contains heartbeat sensor, acceleration sensor and can collect vital data continuously. Sony SmartEyeglass\cite{Sony} is a eyeglass type computer and can collect acceleration and luminance data. Hitoe\cite{hitoe} is a T-shirt type wearable sensor NTT and Toray develop and can collect Electrocardiograph(ECG) and 3-axis acceleration data by wearing hitoe shirt. Vital data is forwarded to a smart phone via a hitoe transmitter on breast. Regarding to acceleration data, horizon direction is X-axis, longitudinal direction is Y-axis and vertical direction is Z-axis (Figure 1).

 \begin{figure}[tb]
 \begin{center}
  \includegraphics[width=64mm]{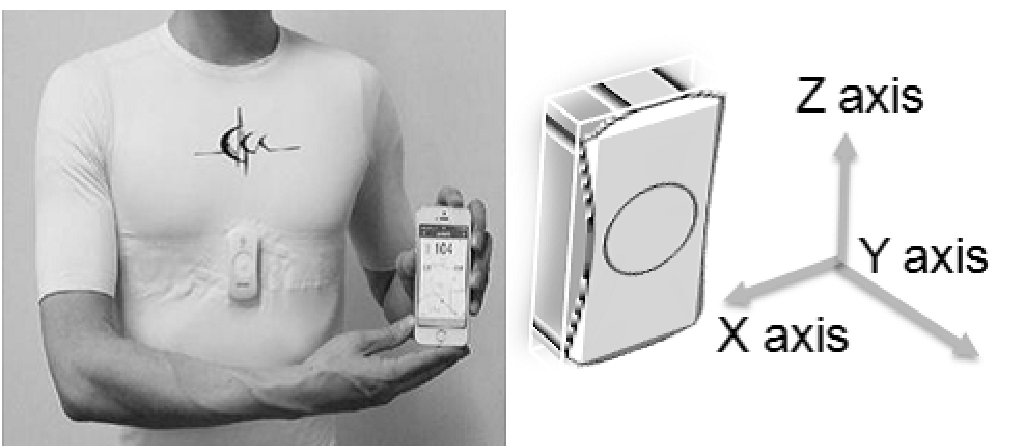}
 \end{center}
 \caption{hitoe and axes of acceleration data}
 \end{figure}

There are tools to estimate posture by analyzing acceleration acquired from these sensors (e.g., NTT Docomo provides hitoe SDK \cite{SDK} for hitoe data analysis).  However, acceleration of vehicles is added to acceleration of wearable sensor in vehicles, it is vague to estimate drivers' posture accurately by tools of wearable sensors. We need to verify posture estimation in vehicles.

%% file: section3E.tex
\section{Ideas of methods}
For drivers' posture estimation in vehicles, we need to consider two things. The first is that acceleration data of wearable sensor includes acceleration of vehicle. The second is that considering safety management of vehicles, specific dangerous posture such as picking up things during driving needs to be detected.  

Based on above, there is a method to estimate drivers' posture by subtracting acceleration of vehicle from the one of wearable sensor. Method 1: Subtracting acceleration of smart phone from the one of hitoe. This method subtracts acceleration of smart phone which a driver holds from acceleration of hitoe which a driver wears with same timestamp. This method may extract acceleration of drivers themselves, but feasibility of subtracting acceleration is concerned such as difference of two sensors accuracy.

Then, there is a method to detect acceleration patterns of specific posture such as picking up things by analyzing hitoe acceleration data. Method 2: Detecting specific posture patterns from hitoe acceleration data. Because drivers of bus and taxi are instructed to avoid sudden departure, acceleration of X-axis and Y-axis is not so huge. On the other hand, because picking up things has postural changes such as forward bending, acceleration of Y-axis and Z-axis of hitoe changes much due to the influence of gravity. Thus, by focusing change of acceleration, we may detect specific dangerous posture with specified threshold values.

%% file: section4E.tex
\section{Verification of methods}
We verified method 1 and 2 based on actual acceleration data using hitoe and smart phone (Sony Xperia).

For method 1 verification, we put hitoe and smart phone on a hand cart and moved it as start, stop, go straight, turn left or right to confirm feasibility of subtracting acceleration of smart phone. Figure 2 shows data of Y-axis acceleration. From Fig.2, acceleration data of hitoe and smart phone are similar but a difference between them does not reach zero. When we reduce high frequency noise by low pass filter, the difference remains. This is because there are differences of reaction and accuracy of two sensors, and method 1 has some problems.

 \begin{figure}[tb]
 \begin{center}
  \includegraphics[width=84mm]{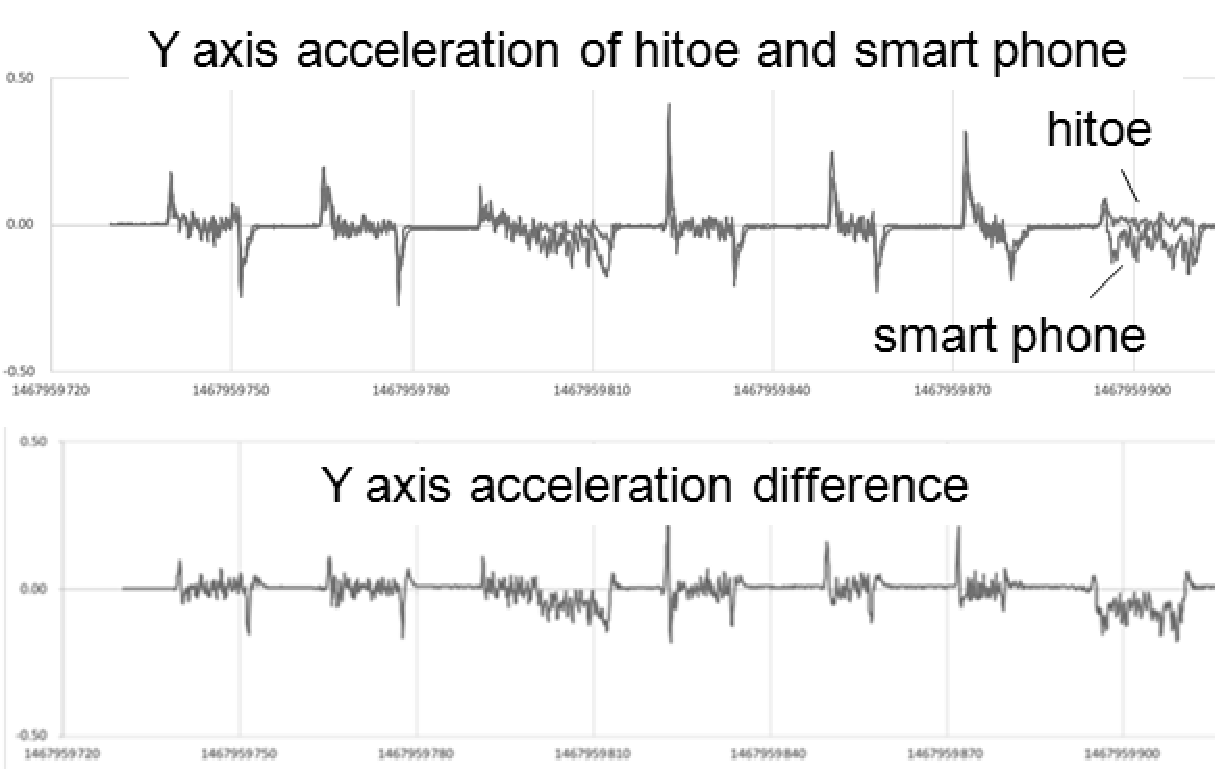}
 \end{center}
 \caption{Y-axis acceleration data of hitoe and smart phone during movement on a hand cart}
 \end{figure}

For method 2 verification, hitoe wearer got on a regular bus and changed posture while seating like picking up things to confirm feasibility of posture detection analysis from hitoe acceleration data. Figure 3 shows acceleration data on a regular bus. From Fig.3, acceleration change from bus is mainly slight changes of X-axis and Y-axis and is not so large. On the other hand, as for picking up things while sitting, it can be seen that the acceleration change on the Y axis is large since the inclination of the body changes. Thus, by focusing acceleration change of Y-axis, we can detect posture change like picking up things. Specifically, we set -0.34G as threshold value of Y-axis acceleration (it is 20 degrees angle), we can detect picking up behavior in a regular bus.

However, when we set threshold value of Y-axis, there is a concern of false detection during steep slope. Therefore, we actually tested on the Mt. Fuji climbing bus which is known for a steep gradient (the slope is about 20\%). As a result, there is no false detection because acceleration of Y-axis does not reach -0.34G even on Mt. Fuji slope.

 \begin{figure}[tb]
 \begin{center}
  \includegraphics[width=84mm]{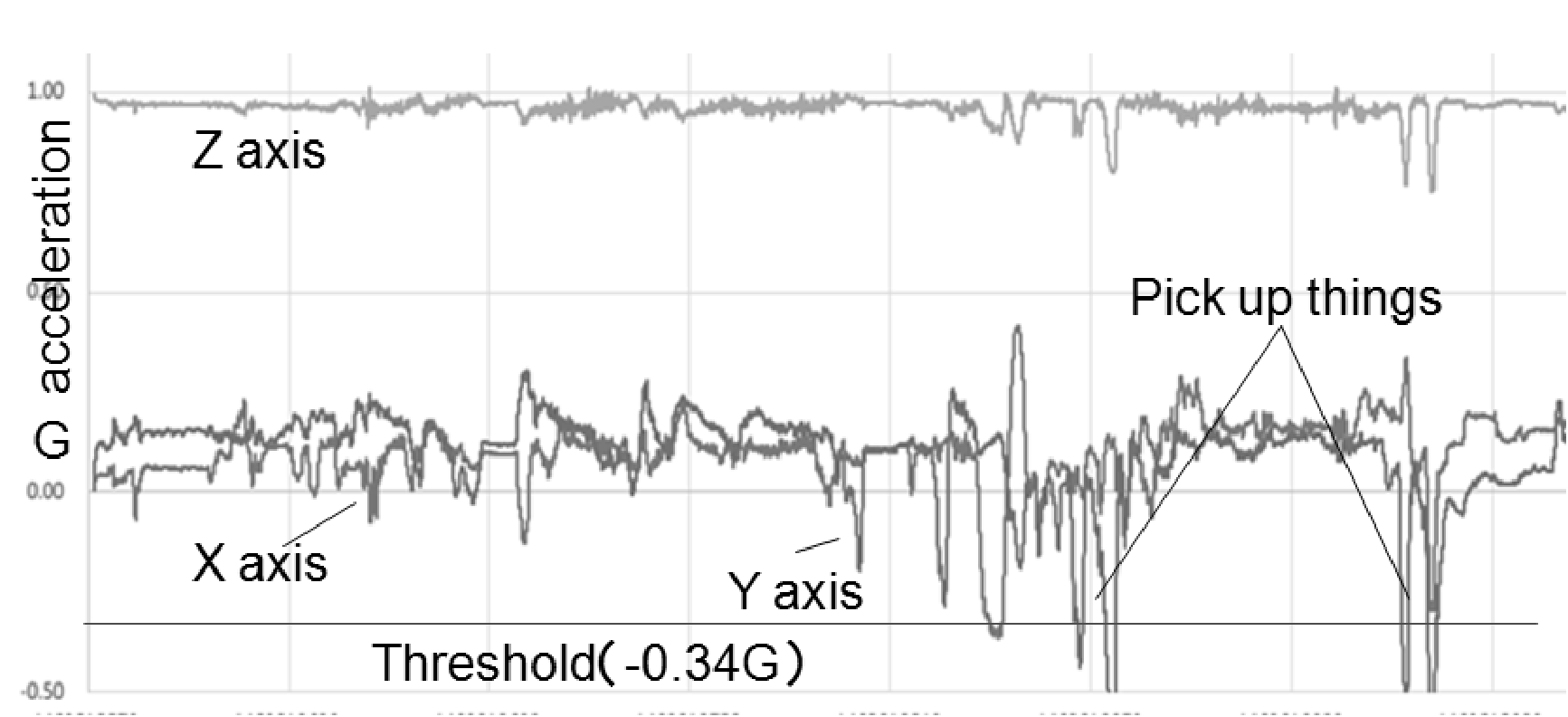}
 \end{center}
 \caption{hitoe acceleration data on a regular bus}
 \end{figure}

%% file: section5E.tex
\section{Sample application}
Based on method 2 verified in the previous section, we implemented a sample application which detects posture changes like picking up things in vehicles. Figure 4(a) shows an outline of sample application processing. The sample application firstly reduces noise for acceleration data from hitoe. Then it shows posture image using processed acceleration data and threshold data.

We tested the implemented sample application in a regular bus in Musashino-shi, Japan, and confirmed a change of posture image when a user picked up a thing (Figure 4(b)).

 \begin{figure}[tb]
 \begin{center}
  \includegraphics[width=84mm]{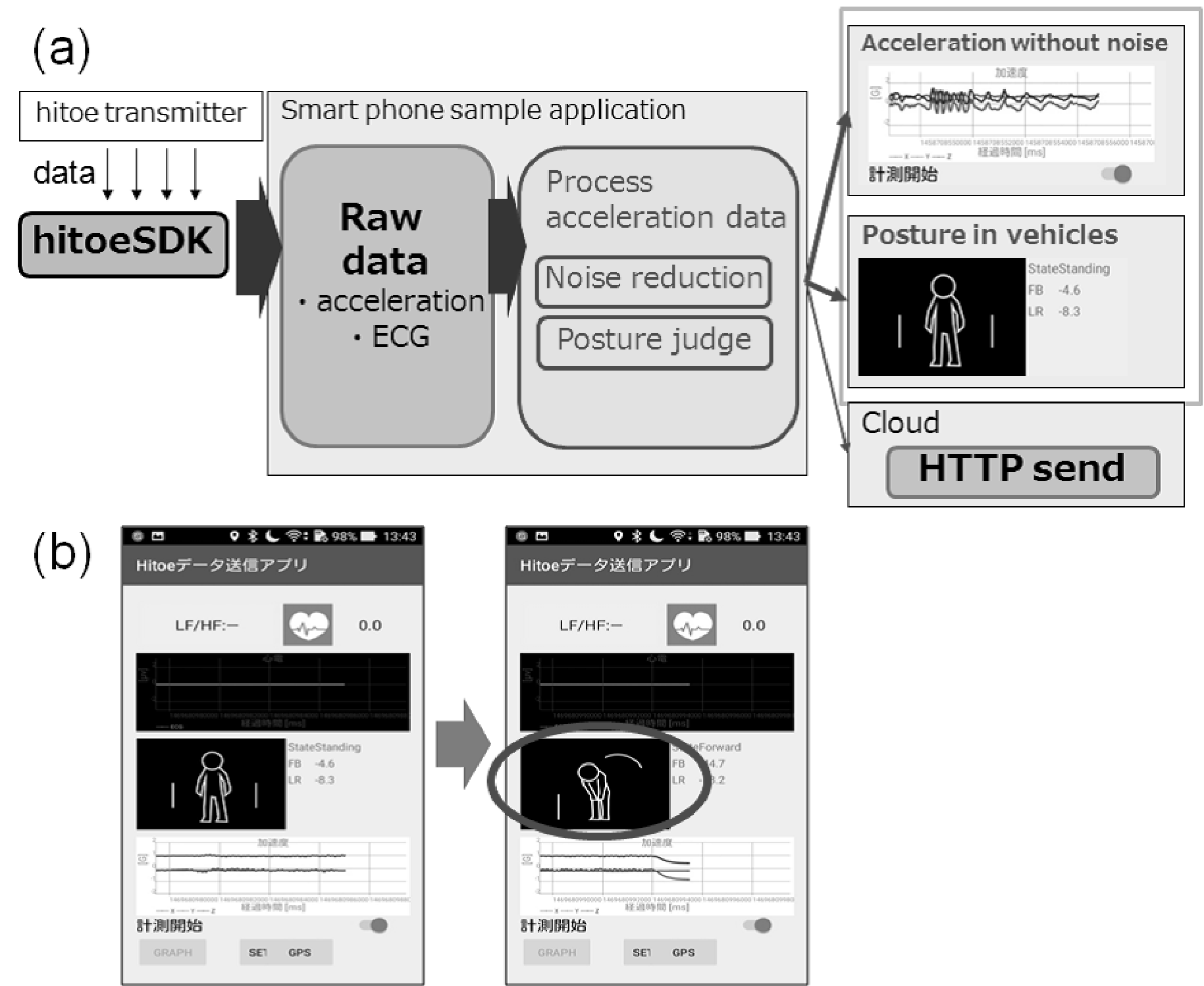}
 \end{center}
 \caption{(a) Outline of sample application processingD(b) GUI images of sample application}
 \end{figure}

A sample application not only shows postures but also sends acceleration data and vital data such as heart rate acquired from hitoe to a cloud, and a cloud does advanced analysis based on cloud technologies such as \cite{Yamato2}-\cite{JIP}. Fatigue level is calculated using RRI change (e.g., \cite{EV}) and relaxation level is calculated with cardiac vagal index (CVI) \cite{CVI}. When machine learning technologies are used, GPU or baremetal power can be used to analyze \cite{JIP2}\cite{SOCA}. Based on analyzed results of cloud, actions of substitute drivers assignments or emergent alerts are conducted using Web services \cite{ICIN}-\cite{ICWS} or service coordination technologies \cite{CCNC}-\cite{WTC}.

%% file: section6E.tex
\section{Conclusions}
This paper studied methods to estimate postures of drivers on vehicles using wearable acceleration sensor hitoe and conducted field tests. The method to subtract vehicle acceleration using hitoe and smart phone has problems of accuracy differences between them. On the other hand, posture changes such as picking things can be detected by threshold judgement of acceleration. Based on the results, we implemented a sample application which shows posture on vehicles and confirmed feasibility of posture estimation.

In the future, we plan a field trial with bus companies using driving posture monitoring and fatigue level monitoring based on hitoe ECG data.